\documentclass[aps,twocolumn,pra,superscript,floatfix,superscriptaddress,showpacs]{revtex4-1}
\usepackage{amssymb}

\usepackage{epsf}
\usepackage{epsfig}
\usepackage{graphicx}
\usepackage{dcolumn}
\usepackage{braket}
\usepackage{bm}
\usepackage{amsfonts}
\usepackage{amsmath}
\usepackage{amssymb}
\usepackage{color,soul}
\usepackage{wasysym}
\usepackage{mathrsfs}
\usepackage{natbib}
\usepackage{float}
\usepackage{multirow}
\usepackage[colorlinks=true,%
            linkcolor=blue,%
            urlcolor=blue,%
            citecolor=blue,%
            filecolor=blue,%
            bookmarksopen=true,%
            pdfauthor={J. P. Ramos-Andrade},%
            pdftitle={class_measurement},%
            pdfsubject={class_measurement_Manuscript},%
            pdfpagemode=UseOutlines]{hyperref}

\newcommand{\ve}{\varepsilon}

\newcommand{\bea}{\begin{eqnarray}}
\newcommand{\eea}{\end{eqnarray}}

\newcommand{\up}{\uparrow}
\newcommand{\dn}{\downarrow}

\begin{document}

\title{Kondo effect in a quantum dot embedded between topological superconductors}
\author{G. A. Lara}
\affiliation{Departamento de F\'isica, Universidad de Antofagasta, Av. Angamos 601, Casilla 170, Antofagasta, Chile.}
\email{gustavo.lara@uantof.cl}	

\author{J. P. Ramos-Andrade}
\affiliation{Departamento de F\'isica, Universidad de Antofagasta, Av. Angamos 601, Casilla 170, Antofagasta, Chile.}

\author{D. Zambrano}
\affiliation{Departamento de F\'isica, Universidad T\'ecnica Federico Santa Mar\'ia, Casilla 110 V, Valpara\'iso, Chile}

\author{P. A. Orellana}
\affiliation{Departamento de F\'isica, Universidad T\'ecnica Federico Santa Mar\'ia, Casilla 110 V, Valpara\'iso, Chile}

\begin{abstract}
In this article, we study the quantum transport through a single-level quantum-dot in Kondo regime, coupled to current leads and embedded between two one-dimensional topological superconductors, each hosting Majorana zero modes at their ends. The Kondo effect in the quantum dot is modeled by mean-field finite-$U$ auxiliary bosons approximation and solved by using the non-equilibrium Green's function approach. First, we calculate the density of states of the quantum dot, and then both the current and the differential conductance through the quantum dot in order to characterize the interplay between the Kondo resonance and Majorana zero modes. The results reveal that the presence of Majorana zero modes modifies the Kondo resonance exhibiting an anti-resonance structure in the density of states, leading to obtain spin-resolved behavior of the measurable current and differential conductance. We believe our findings could be helpful to understand the behavior of the Kondo effect in connection with Majorana zero modes.
\end{abstract}

\date{\today}
\maketitle

\section{Introduction}\label{intro}

The Majorana modes are zero-energy states emerging in topological superconductors (TSCs) \cite{alicea2011non,alicea2012}. One of the main features of these so-called Majorana zero modes (MZMs) is that they satisfy non-Abelian statistics, and for this feature, they are considered potential candidates for quantum computation implementations \cite{wilczek2009majorana}, which is why they have attracted a great deal of attention from condensed matter physicists. Among other theoretical systems, one of the most studied is the model proposed by Kitaev, in which MZMs are predicted to be found at both ends of a one-dimensional (1D) chain \cite{kitaev2001}. Moreover, a physical realization of a Kitaev chain has been achieved in a 1D semiconductor structure, when proximitized by a superconductor and in the presence of a magnetic field, and confirmed via electronic measurements \cite{lutchyn2018majorana}. Thus, in the quest for evidence of the presence of Majorana modes, the measurement of transport quantities arises as a promising tool. In this context, Mourik et al. reported for the first time zero-bias anomalies in the conductance as a Majorana mode signature in the system \cite{mourik2012signatures}. Starting from this point, the advance in experimental techniques and the consideration of theoretical proposals have led, among others, to the use of quantum dots (QDs) in interplay with MZMs \cite{albrecht2016exponential,deng2016majorana,das2012zero}, since they own regular fermionic states. In this scenario, both theoretical predictions and experimental measurements have shown zero-bias peaks in a QD connected with MZMs \cite{liu2011}, regardless the QD's energy level \cite{vernek2014,RamosAndrade2018}. Therefore, this phenomenon was ascribed as a MZM-leakage into the QD \cite{vernek2014,deng2016majorana}.      

On the other hand, the incorporation of QDs into systems hosting MZMs has posed the question of how do quantum correlation effects might influence eventual measurements \cite{zambrano2018,RamosAndrade2019,RamosAndrade2020}. The Kondo effect in single occupation QDs shows itself as conductance peaks close to zero-energy \cite{Kouwenhoven_2001}. Then, it is natural to wonder how will this phenomenon (non-topological) compete with the one caused by the presence of MZM (topological) \cite{Golub2011kondo}. Previous works addressing this question have concluded that the zero-energy peak depends on the temperature, that the low-temperature transport properties are indeed modified, and that the Kondo fixed point becomes unstable 
\cite{Gao2016Kondo,Lee2013kondo,YANG2020Kondo,gorski2017kondo,Cheng2014interplay,Weymann2020MajoranaKondo,PhysRevB.91.081405,liu2020topological}. Also, the MZM-leakage phenomenon has been studied in interacting QD \cite{ruiz2015}. In recent work, the authors showed that the Kondo effect in a QD is robust against coupling with MZMs belonging to different superconductor-semiconductor wires with opposite spin polarizations \cite{Silva2020Robustness}. The diversity of conclusions evidences that the interplay between Kondo and MZMs is not yet fully understood.

Our interest in the present work is to study the interference phenomena in a single energy-level QD device, considering the presence of MZMs provided by 1D TSCs. 
In the system under study, the QD is embedded between both MZMs and metallic conductors. We describe the system using an effective low-energy Hamiltonian, where 1D TSCs and the MZMs are modeled using the Kitaev model. The transport properties, such as the QD's density of states and current, have been calculated employing the non-equilibrium Green's function. We considered the mean-field finite-$U$ auxiliary bosons formalism to achieve the Kondo regime in the QD. Although this formalism does not fully describe all the characteristics of the Kondo effect, we emphasize that it does capture its essence, and therefore constitutes a reliable way to treat the Kondo regime in the system. Our results show that Kondo resonance in the QD's density of states is affected due to the coupling with MZMs, showing changes on its amplitude at zero energy depending on the interference phenomena between the MZMs and the coupling regime with the QD. We present the current and differential conductance through the QD regarding the applied bias to emphasize the features caused by the interplay between both Kondo and MZMs phenomena.

This paper is arranged as follows: in Sec.\ \ref{sec:model} we described the model and the method considered, followed by presenting the results and the related discussion in Sec.\ \ref{sec:results}. A summary and final comments are presented in Sec.\ \ref{sec:conclu}.

\section{Model and method}\label{sec:model}

\begin{figure}[htb!]
\centering
\includegraphics[width=1.00\linewidth]{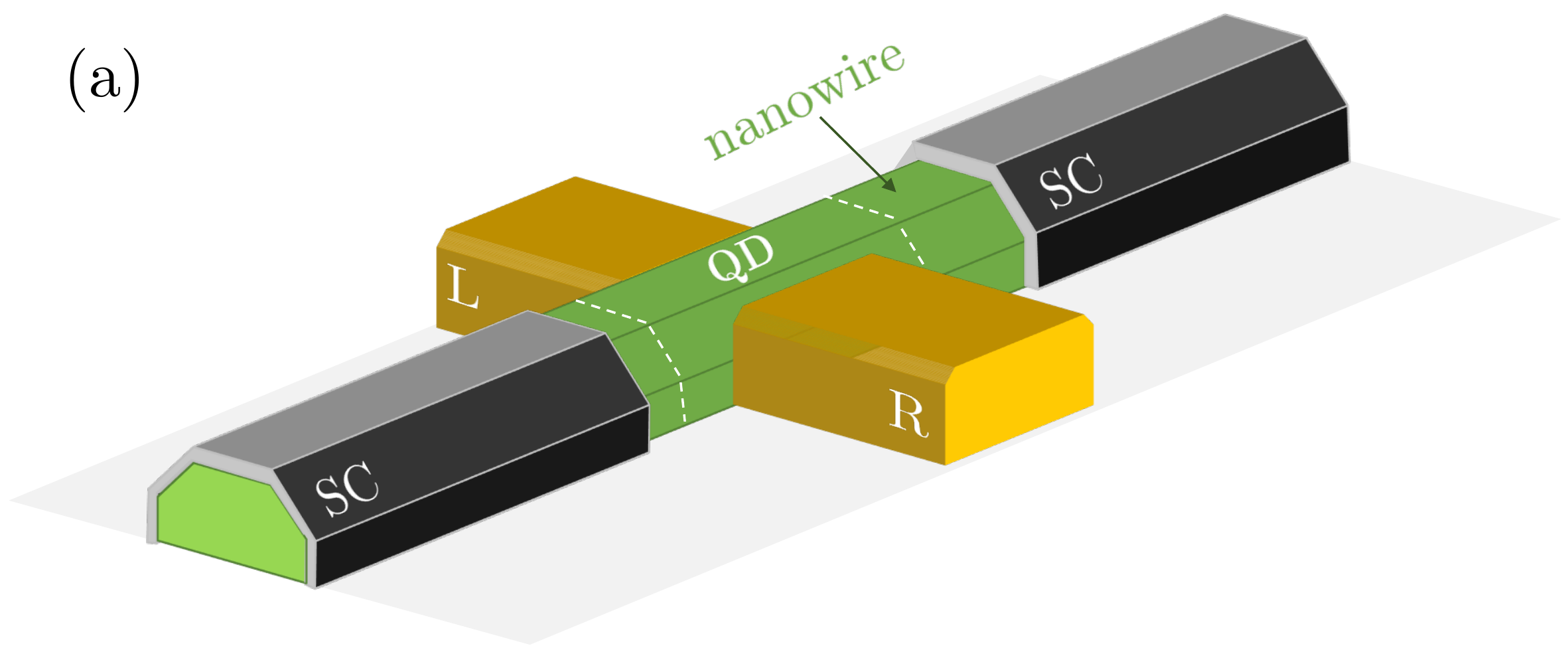}\\
\includegraphics[width=1.00\linewidth]{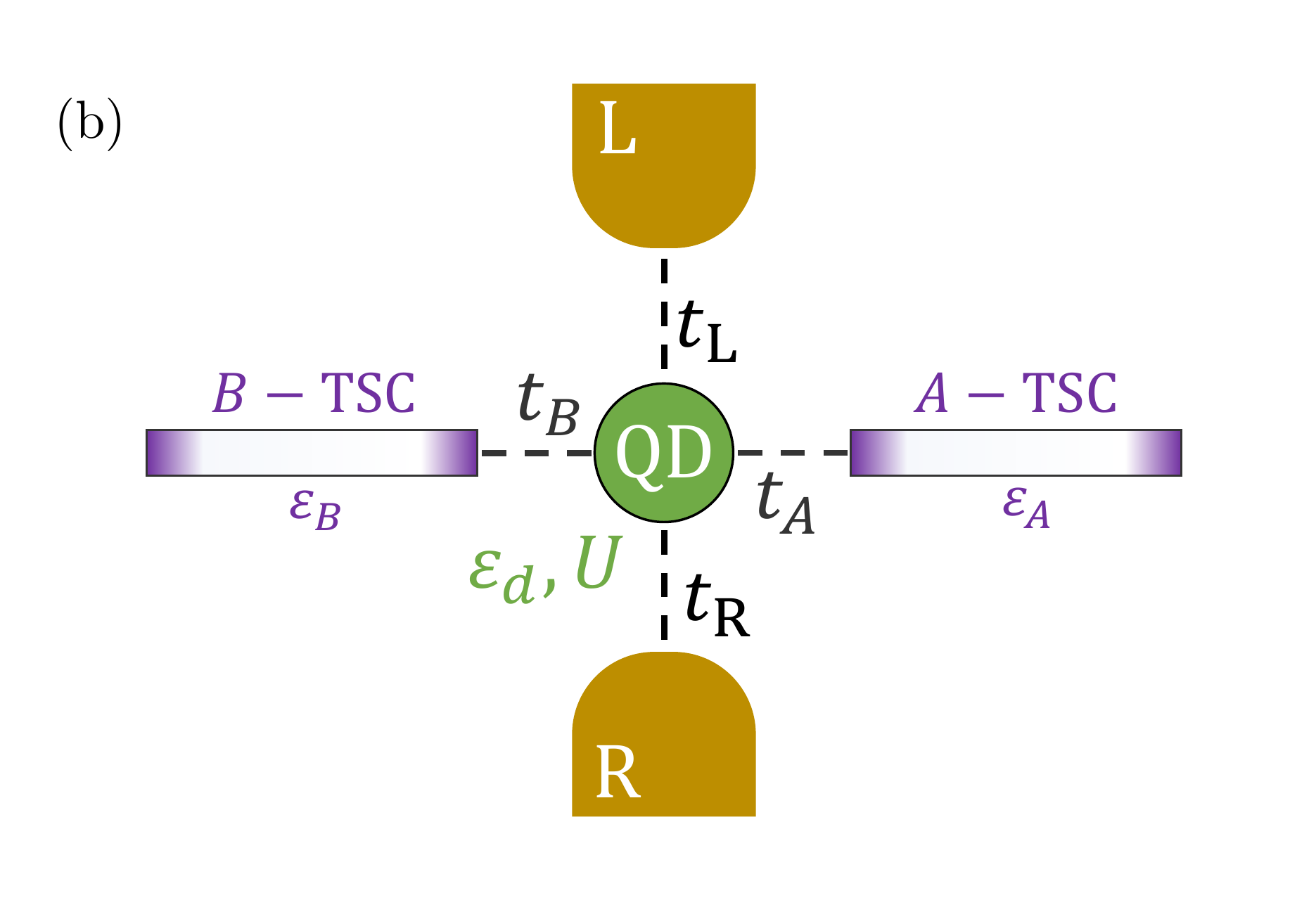}
\caption{\label{Figure_Model} (a) Setup under study: semiconducting nanowire with intrinsic spin-orbit coupling proximitized by superconductors in two zones placed at the ends. Under an appropriated magnetic field, the covered zones corresponds to TSCs, while the unconvered form a QD. The latter is embedded between two regular leads. (b) Schematic view. The QD is connected to the two leads, L and R, via tunneling couplings $t_L$ and $t_R$, respectively. The two TSCs considered, $A$ and $B$, are connected to the QD with tunneling couplings $t_A$ and $t_B$, respectively. The two MZMs placed at the ends of the $A(B)$-TSC (purple zones) are connected between them with a coupling parameter $\varepsilon_{A(B)}$.}\label{fig1}
\end{figure}

We study a system formed by a single-QD embedded between current leads and connected to two TSCs hosting MZMs at their ends. The setup considered is presented in Fig.\ \ref{Figure_Model}, where the different components and the respective couplings between them are described.

We model the system with a Hamiltonian in the form
\begin{equation}\label{Hinicial}
 H = H_{\text{l-d}} + H_{\text{s-d}}\,,
\end{equation}
where the first term describes the leads, QD, and their connections, while the second term describes the TSCs, i.e., the MZMs and the connections between them and with the QD.  

The lead $\alpha$ ($\alpha=L,R$) has a chemical potential $\mu_{\alpha}$, that can be controlled by means of an electric potential. We consider the leads' chemical potential as the energy reference. Then, by applying a voltage $V$ between the leads, $\mu_{L}=-eV/2$ and $\mu_{R}=eV/2$ are obtained. 
In each lead the occupation probability distribution is given by the Fermi function 
$f_{\alpha} (\varepsilon_{k}) = \left[ 1+\exp{\left\{ \beta \left( \varepsilon_{k} +eV_{\alpha} \right) \right\}} \right]^{-1} $, where $\ve_{k}$ is the energy and $\beta=1/\kappa_{\text{B}}T$, being $T$ the temperature and $\kappa_{\text{B}}T$ the Boltzmann constant.

The single-level energy of the QD is $\varepsilon_d$, which splits due to the Zeeman effect. Additional Coulomb energy $U$ is considered in the double occupancy regime to describe the electrostatic interaction between electrons in confinement. 
Then, the Hamiltonian for the subsystem leads-QD is expressed as  
\begin{equation}
\begin{aligned} 
 H_{\text{l-d}} 
 = & 
  \sum_{\alpha,\mathbf{k}_{\alpha},\sigma } 
  \left[ 
       \varepsilon_{\mathbf{k}_{\alpha}} \hat{n}_{\mathbf{k}_{\alpha} \sigma }
 -t_{\alpha} \left( c_{\mathbf{k}_{\alpha}\sigma }^{\dag} f_{0\sigma} 
                   +f_{0\sigma}^{\dag} c_{\mathbf{k}_{\alpha}\sigma }
            \right) 
 \right] 
\\ &
+ \sum_{\sigma } 
 \left( \varepsilon_{d} +\sigma g\mu_{\text{B}} B \right)  \hat{n}_{0\sigma}
 +U \hat{n}_{0\uparrow} \hat{n}_{0\downarrow} \,, 
\end{aligned} 
\end{equation}
where $\hat{n}_{\mathbf{k}_{\alpha},\sigma} =  c_{\mathbf{k}_{\alpha},\sigma }^{\dag} c_{\mathbf{k}_{\alpha},\sigma }$ and 
$\hat{n}_{0\sigma}= f_{0\sigma}^{\dag}f_{0\sigma}$ are the number operators in the lead-$\alpha$ and in the QD, respectively, with $\sigma$ denoting the electronic spins ($\uparrow$ or $\downarrow$). The applied magnetic field corresponds to $B$, $\mu_{\text{B}}$ is the Bohr magneton, and $g$ is the Land\'e factor.

The MZMs quasi-particles are their own anti-quasi-particles. Accordingly, with the Kitaev model, we treat the MZMs operators as a superposition of regular fermionic operators $\tilde{c}_{\nu}$, being $\nu=A,B$ (see Appendix \ref{apenA}).
Each TSC interacts with the QD by transferring a fermionic state between them, coupling a Majorana operator from the TSC with a Majorana operator arising in the QD. Then, following the result shown in Eq.\ (\ref{ham-TSC-QD}), the Hamiltonian $H_{\text{s-d}}$ is given by

\begin{equation} \label{Hsd}
 H_{\text{s-d}} = \! \! \sum_{\nu=A,B}
 \left[ 
 \varepsilon_{\nu} \tilde{c}_{\nu}^{\dag} \tilde{c}_{\nu}
 + \left( \tilde{t}_{\nu} f_{0\downarrow}^{\dag} 
         +\tilde{t}_{\nu}^* f_{0\downarrow} \right) 
 \left( \tilde{c}_{\nu}^{\dag}  -\tilde{c}_{\nu} \right)
\right] ,
\end{equation}
where $\varepsilon_{\nu}\propto \exp{(-L_{\nu}/\xi_{\nu})}$ and $t_{\nu} = |t_{\nu}| e^{i\theta_{\nu}}$, with $|t_{\nu}|$ the coupling amplitude between the $\nu$-TSC and the QD, and $\theta_{\nu}$ the phase of electrons in the $\nu$-TSC. Note that since $L_{\nu}$ is the wire's length, $\ve_{\nu}\to 0$ corresponds to the long wire limit, and nonvanishing $\ve_{\nu}$ to the short wire limit.

We considered the Coulomb interaction in the QD using auxiliary bosons, following the approach show in the Appendix \ref{apenB}. 
Then, within the mean field approximation, the Hamiltonian is given by
\begin{equation}
 \begin{aligned} 
  H_{\text{mfa}} = &
 \sum_{\alpha, \mathbf{k}_{\alpha},\sigma }   
 \Big\{ \varepsilon_{k_{\alpha}} \hat{n}_{\mathbf{k}_{\alpha} \sigma } 
       -\tilde{t}_{\alpha \sigma} 
       \left( c_{k_{\alpha} \sigma }^{\dag}  f_{0\sigma} 
             +f_{0\sigma}^{\dag} c_{k_{\alpha}\sigma }
            \right)
\Big\} 
\\ & 
+\sum_{\sigma} \widetilde{\varepsilon}_{\sigma} \hat{n}_{0\,\sigma}
 + \sum_{\nu}  
 \varepsilon_{\nu} \tilde{c}_{\nu}^{\dag} \tilde{c}_{\nu}
 \\ &  
 + \sum_{\nu}  
 \left( \tilde{t}_{\nu} f_{0\downarrow}^{\dag} 
         +\tilde{t}_{\nu}^* f_{0\downarrow} \right) 
 \left( \tilde{c}_{\nu}^{\dag}  -\tilde{c}_{\nu} \right)
 \\ & 
+\lambda_{0}^{(1)}
 \left(  e^2 + {p}_{\uparrow }^2  
 + \hat{p}_{\downarrow}^2 +d^2  - 1 \right) 
\\ & 
 -\sum_{\sigma } \lambda_{0\, \sigma}^{(2)}
 \left( {p}_{ \sigma }^2 +d^2 \right) +U d^2\,,
 \end{aligned}
\end{equation}
where $\tilde{t}_{\alpha \sigma} = t_{\alpha} Z_{\sigma}$, 
$\tilde{t}_{\nu} = t_{\nu} Z_{\downarrow} $, and 
$
\widetilde{\varepsilon}_{\sigma} 
= 
 \varepsilon_{0} +\sigma g\mu_B B  + \lambda_{0\, \sigma}^{(2)}$. The parameters $ e$, $ d$, $ p_{\sigma}$, $ \lambda^{(1)}_{0}$, and $\lambda^{(2)}_{0\, \sigma}$,  
are determined by minimizing the ground state energy of the Hamiltonian $H_{\text{mfa}}$. 


To obtain the physical quantities of interest, we employ the Keldysh Green's functions (GFs) formalism for stationary states out-of-equilibrium, defining the regular GFs as
\[ 
\begin{aligned} 
 {\cal G}_{p 0}^{\sigma} (t,t') =  & 
 \frac{-i}{\hbar} 
 \Big\langle T_K c_{p \sigma}(t) f_{0  \sigma}^{\dag}(t') \Big\rangle\,,
\\ 
 {\cal G}_{0 0}^{\sigma} (t,t') =  & 
 \frac{-i}{\hbar} 
 \Big\langle T_K f_{0 \sigma}(t) f_{0  \sigma}^{\dag}(t') \Big\rangle\,,
 \end{aligned} 
\] 
and the anomalous GFs as
\[ 
\begin{aligned} 
 {\cal F}_{\nu 0} (t,t') =  & 
 \frac{-i}{\hbar} 
 \Big\langle T_K \tilde{c}_{\nu \downarrow}^{\dag} (t) f_{0\downarrow}^{\dag}(t') \Big\rangle\,,
\\ 
 {\cal F}_{00} (t,t') =  & 
 \frac{-i}{\hbar} 
 \Big\langle T_K f_{0\downarrow}^{\dag} (t) f_{0\downarrow}^{\dag}(t') \Big\rangle\,.
 \end{aligned} 
\] 
Note that the latter are spin-resolved since only spin-down electrons are coupled with MZMs \cite{ruiz2015}. Applying Langreth's rules to the set of integral equations of motion, the following results are obtained for GFs in the energy domain
\begin{eqnarray}
  G_{00}^{\uparrow,a}
 &= &
 \frac{\varepsilon -\widetilde{\varepsilon}_{\uparrow} +i\Gamma_{T \uparrow }}
      {\left(\varepsilon-\widetilde{\varepsilon}_{\uparrow}\right)^2 
       + \Gamma_{T\uparrow}^2  }\,, \\
G_{0 0}^{\uparrow ,<}
 &=&
 \frac{i \sum_{\alpha} \Gamma_{\alpha} Z_{\uparrow}
                     2 f(\varepsilon + eV_{\alpha}) }
      {\left(\varepsilon-\widetilde{\varepsilon}_{\uparrow}\right)^2 
       + \Gamma_{T\uparrow}^2  }\,,\\
   G_{00}^{\downarrow,a} 
  & = &
\frac{1}{D} 
\left[ 
 \varepsilon_n \left( \varepsilon_{p}^2 +\Gamma_{T\downarrow}^2 \right)
 -\varepsilon_p \left|\widetilde{R}_{-1}\right|^2
  +i\Gamma_{T\downarrow} D_{p}^2
 \right],\\
  F_{00}^{a} 
  &=&
 \frac{ \widetilde{R}_{-1} }{D}
\left[ 
      \varepsilon_p (\varepsilon_n +\varepsilon_p) -D_{p}^2  
      +i\Gamma_{T\downarrow} \left( \varepsilon_n +\varepsilon_p \right)
\right],\\
G_{0 0}^{\downarrow ,<} 
  &=& 
\frac{i}{D} \left[   
      \Gamma_{+z}
      \left( \varepsilon_{p}^2 +\Gamma_{T\downarrow}^2 \right) 
     +\Gamma_{-z} \left| \widetilde{R}_{-1}\right|^2
   \right]\,,\\
F_{B}^{<}  
&=&    \\
&&i \frac{\widetilde{R}_{-1}}{D}  
\left[   
      \left( \Gamma_{+z} \, \varepsilon_p 
            +\Gamma_{-z} \, \varepsilon_n  
      \right)
      -i\Gamma_{T\downarrow} 
       \left( \Gamma_{+z} -\Gamma_{-z} \right)
 \right] \,,\nonumber \\
F_{00}^{<}  
&=&
F_{B}^{<} 
-
\delta(\hbar\omega+\widetilde{\varepsilon}_{\downarrow})
\int\limits_{-\infty}^{\infty} 
d(\hbar\omega') \, F_{B}^{<} (\omega')\,,
\end{eqnarray}
where the following quantities have been defined:
\begin{eqnarray}
\Gamma_{\pm z} &=& \sum_{\alpha} \Gamma_{\alpha} Z_{\downarrow} 2f(\varepsilon\pm eV_{\alpha} ) 
\,; 
\Gamma_{T\sigma} = \sum_{\alpha} \Gamma_{\alpha} Z_{\downarrow}\,,\\ 
\widetilde{R}_{n} &=& \sum_{\nu} |\tilde{t}_{\nu}|^{2} e^{n2i\theta_{\nu}} 
{\cal P} \left\{ \frac{1}{\varepsilon-\varepsilon_{\nu} } + \frac{1}{\varepsilon+\varepsilon_{\nu} } \right\}\,, \\
\varepsilon_{\left(\substack{p\\n}\right)} 
&=&
\varepsilon \pm \widetilde{\varepsilon}_{\downarrow} -\widetilde{R}_0
\,, \, 
 D_{p}^2 = \varepsilon_{p}^2 +\Gamma_{T\downarrow}^2 
   + \left|\widetilde{R}_{-1}\right|^2\,,\\
E_{\pm} &=& \widetilde{R}_0  \pm \sqrt{\widetilde{\varepsilon}_{\downarrow} +\left| \tilde{R}_{-1} \right|^2  } \,,\\
    D &=& \left| \left( \varepsilon -E_{+} -i\Gamma_{T\downarrow}\right) 
               \left( \varepsilon -E_{-}  -i\Gamma_{T\downarrow} \right)
       \right|^2 \,.
\end{eqnarray}
Here $\Gamma_{\alpha}=\pi\rho_{\alpha}|t_{\alpha}|^{2}$ correspond to the lead-QD coupling parameter in the wideband limit.

\subsection*{Density of states and electric current}

The spin-resolved local density of states in the QD is given by $\rho_{0\sigma}(\varepsilon)=(1/\pi)\Im{\left\{ G_{00}^{\sigma,a} (\varepsilon) \right\} }$. According with the expressions above, for spin $\sigma=\uparrow$ is given by
\begin{equation}\label{rhoup}
\rho_{0\uparrow}(\varepsilon)  = \frac{1}{\pi\Gamma_{T\uparrow}} 
\frac{\Gamma_{T\uparrow}^2}{\left( \varepsilon -\widetilde{\varepsilon}_{\uparrow} \right)^2 +\Gamma_{T\uparrow}^2 }
\,, 
\end{equation}
while for spin $\sigma=\downarrow$, is given by
\begin{equation}
\rho_{0\downarrow} (\varepsilon)
 =
 \frac{\Gamma_{T\downarrow}}{2\pi }
 \left[ 
 \frac{  
        1+\frac{\widetilde{\varepsilon}_{\downarrow} }
                     {\sqrt{ \widetilde{\varepsilon}_{\downarrow}^2 
                             +\left|\widetilde{R}_{-1}\right|^2 } 
                     }
    }
    { \left( \varepsilon -E_{+} \right)^2 +\Gamma_{T\downarrow}^2 }
+
 \frac{ 
        1-\frac{\widetilde{\varepsilon}_{\downarrow} }
              {\sqrt{ \widetilde{\varepsilon}_{\downarrow}^2 
                     +\left|\widetilde{R}_{-1}\right|^2 } 
              }
    }
    { \left( \varepsilon -E_{-} \right)^2 +\Gamma_{T\downarrow}^2 }
\right] 
.
\end{equation}

The spin-resolved electric current, from the QD to the left lead, is obtained from the following expression
\begin{equation}
 I_{\sigma}
 =
-\frac{2e}{h} \Gamma_{L} Z_{\sigma} 
 \int\limits_{-\infty}^{\infty} d\varepsilon 
\Im{\left\{ G_{00}^{\sigma ,<} 
           -2f(\varepsilon +\tfrac{eV_L}{2}) G_{00}^{\sigma ,a}  
 \right\}} \,.
\end{equation}
Then, for spin $\sigma=\uparrow$ is
\begin{equation}
I_{\uparrow}(V) 
=
\frac{e}{h} \Gamma_{T\uparrow} \int\limits_{-\infty}^{\infty} 
\frac{d\varepsilon \, \left[ f(\varepsilon -\tfrac{eV}{2}) -f(\varepsilon+\tfrac{eV}{2}) \right]} 
{\left( \varepsilon -\widetilde{\varepsilon}_{\uparrow} \right)^2 +\Gamma_{T\uparrow}^2 }\,,
\end{equation}  
while for spin $\sigma=\downarrow$ is
\begin{equation}
I_{\downarrow}(V) 
=
\frac{e}{h} \Gamma_{T\downarrow} \int\limits_{-\infty}^{\infty} 
\frac{d\varepsilon \, D_{p}^2}{D} \left[ f(\varepsilon -\tfrac{eV}{2}) -f(\varepsilon+\tfrac{eV}{2}) \right]\,.
\end{equation}

\section{Results}\label{sec:results}

We present the numerical results obtained and the related discussions throughout this section. We perform all the calculations using symmetric lead-QD coupling parameters 
$\Gamma=\Gamma_{L}=\Gamma_{R}$, used as the energy unit of the system, with a large wideband $W=2\cdot 10^3 \, \Gamma $, and a low enough temperature $k_BT=10^{-9} \, \Gamma$.
Besides, since a magnetic field $B$ is present in physical realizations of Kitaev's model, we considered it such as $g\mu_B B=1\cdot 10^{-4}$ (in units of $\Gamma$) and, since the QD energy level can be controlled by means of a gate voltage, we use $\varepsilon_d=V_{g}$, where $V_{g}$ is the applied gate voltage.  
To consider the existence of the Kondo effect, we look at the existence of a resonance peak near the Fermi level and its half-width is associated with the respective Kondo temperature. 

\begin{figure}[htbp]
\centering
\includegraphics[width=1.00\linewidth]{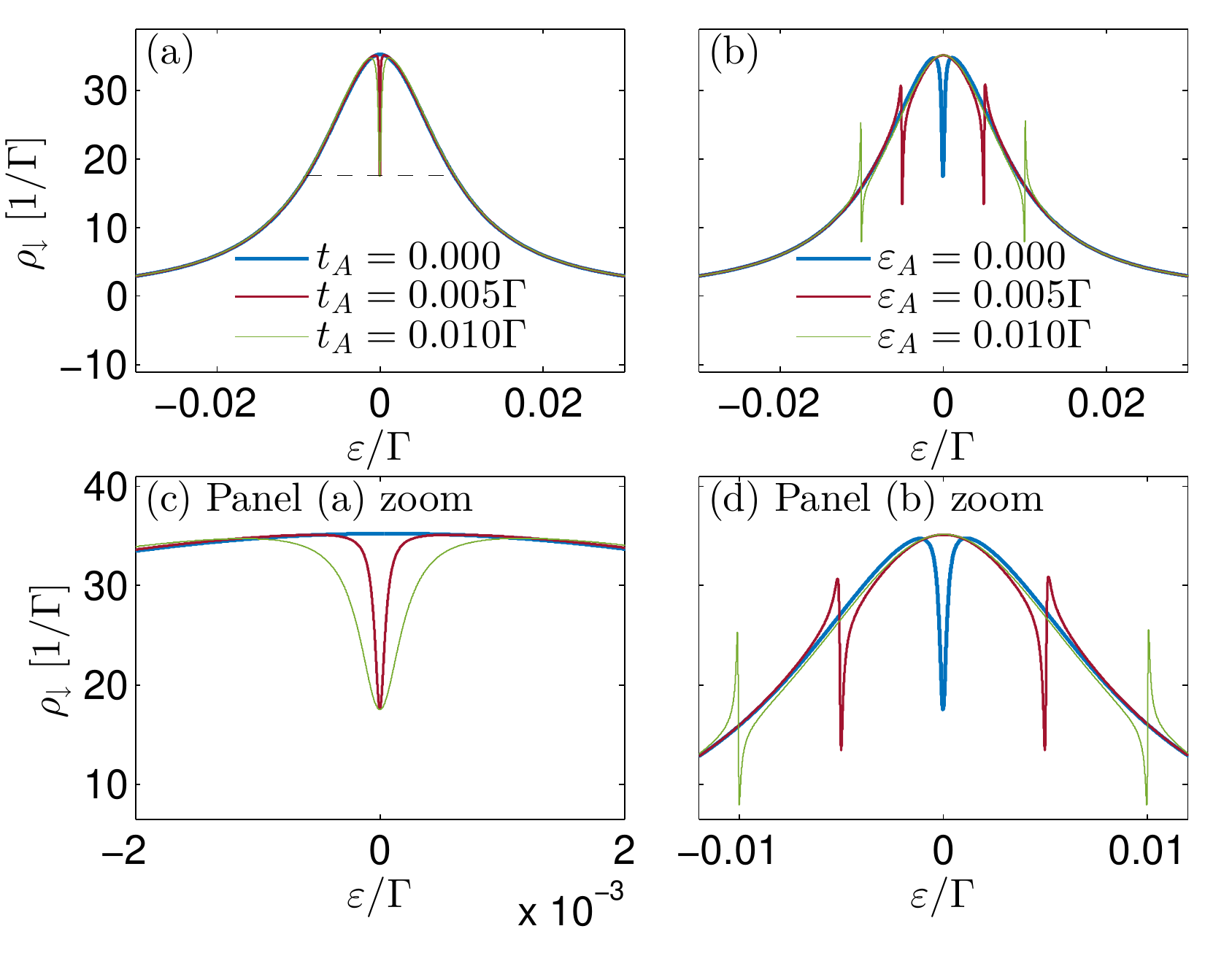}
\caption{QD local density of states $\rho_{\dn}$ as function of the energy $\varepsilon$ for fixed $U=-10V_g$, $V_g=-10\Gamma$, and $t_B = \varepsilon_{B} = \theta_A = \theta_B = 0$. Left panels [(a) \& (c)]: $\varepsilon_A=0$ and different $t_{A}$ values: blue for $t_{A} = 0$, red for $t_{A} = 0.005\Gamma$ and green for $t_{A} = 0.010\Gamma$. The black dashed line represents the FWHM $0.009\Gamma$. Panel (c) is a zoom of panel (a). Right panels [(b) \& (d)]: $t_A=0.010\Gamma$ and different $\varepsilon_{A}$ values: blue for $\varepsilon_{A} = 0$, red for $\varepsilon_{A} = 0.005\Gamma$ and green for $\varepsilon_{A} = 0.010\Gamma$. Panel (d) is a zoom of panel (b).}
\label{rhod}
\end{figure}

\begin{figure}[htbp]
\centering
\includegraphics[width=1.00\linewidth]{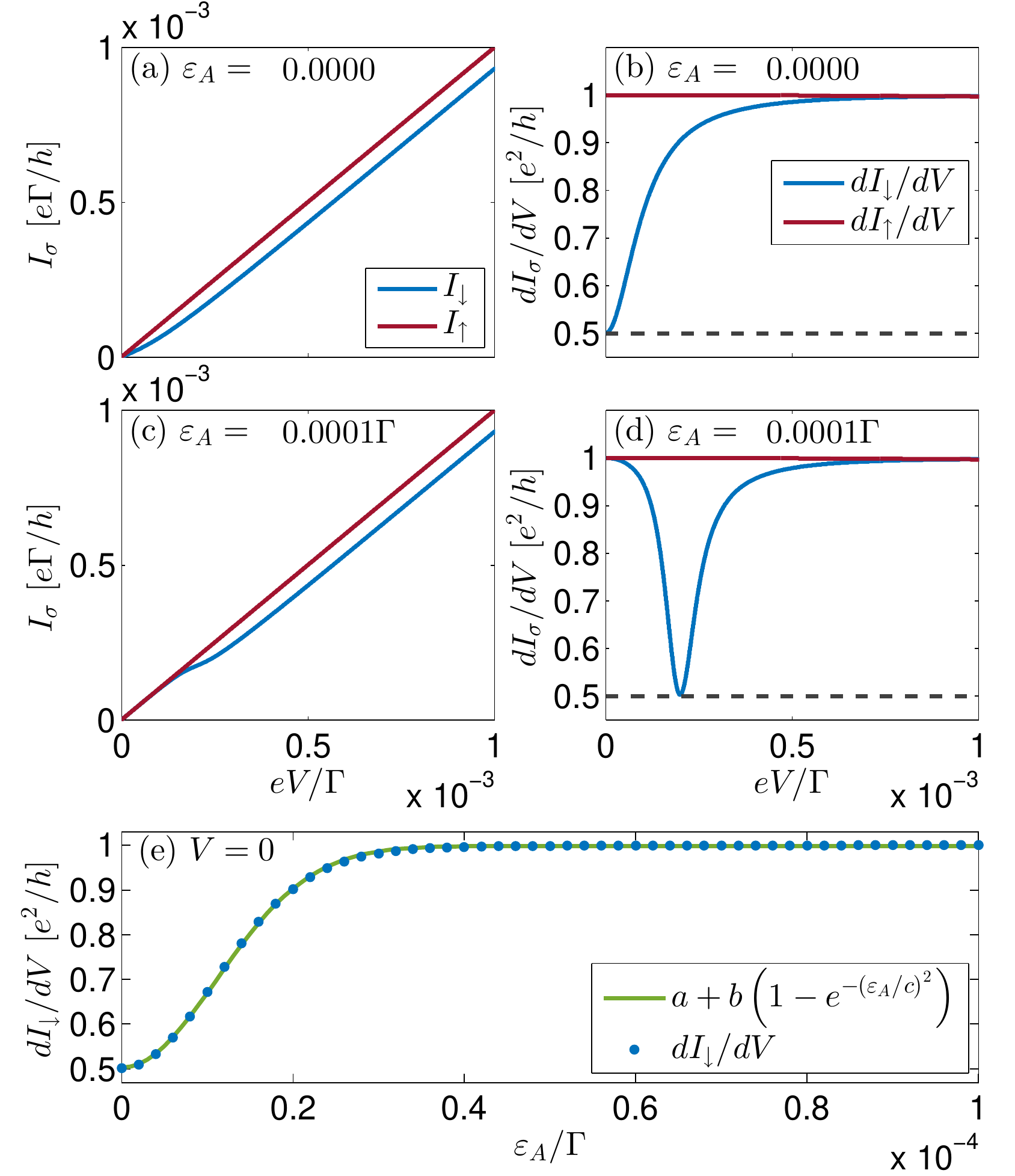}
\caption{Spin dependent current $I_\sigma$ and differential conductance $dI_\sigma/dV$ as a function of the bias voltage $V$ for fixed $U=-10V_g$, $V_g=-10\Gamma$, $t_A=0.005\Gamma$ and $\theta_{A}=0$. Panels (a)-(b) are for $\varepsilon_A=0$; and panels (c)-(d) are for $\varepsilon_A=0.0001\Gamma$. The black dashed lines in the right panels represent the half-integer conductance. Panel (e) displays the zero-bias differential conductance as a function of $\varepsilon_A$ for spin $\sigma=\,\dn$. The fit parameters are: $a=0.5029$, $b=0.4955$ and $c=1.5590\times 10^{-5}$. All panels are for the single TSC case ($t_B=0$, $\varepsilon_B=0$ and $\theta_B=0$).}
\label{IdGvsV_1}
\end{figure}

We start by addressing the case of a single TSC, fixing $t_{B}=0$. In Fig.\ \ref{rhod} we present the QD's local density of states for the spin $\sigma=\,\dn$ component as a function of the energy. In the left panels [(a) and (c)] we considered the TSC in the long-wire limit ($\varepsilon_{A}=0$). For $t_{A}=0$ (blue line), the QD is isolated, and then a resonance around zero energy point is observed, according to the Kondo regime of the system. Whenever $t_{A}\neq 0$ (red and green lines), the resonance is affected by the connection to the MZM, since an interference profile shows a dip of width $\propto t_{A}^{2}$. In the right panels [(b) and (d)], we used a fixed $t_{A}=0.01\Gamma$, and the short-wire limit is considered. We observed that the single dip displayed at zero energy for the long-wire limit (blue line) splits into two Fano-like lines placed at energies $\pm \varepsilon_{A}$ (red and green lines). The QD's local density of states for the spin $\sigma=\,\up$ component is mostly unaffected by the coupling of the MZMs (not shown). The behavior regarding the energy, in this case, corresponds to a Lorentzian line shape according to Eq.\ (\ref{rhoup}) and it is similar to the one presented for the spin $\sigma=\,\dn$ component for $t_{A}=0$. At this point, for the single TSC case, it is clear that the coupling of MZMs, in either the long and short wire limit, influences the Kondo resonance by means of interference phenomena between them. It is remarkable how the central peak is suppressed up to a half-maximum value for the long-wire limit, while in the short-wire limit the maximum value is restored, giving place to side Fano resonances.

Other measurable physical quantities, such as the current, can give us additional information about the coupling between the MZMs and the QD.
In Fig.\ \ref{IdGvsV_1} we present the current [panels (a) and (c)] and the differential conductance across the QD as functions of bias voltage [panels (b) and (d)]. For the spin $\sigma=\,\up$ component, the current exhibits a uniform slope behavior, and as a consequence, the differential conductance is observed as approximately constant, regardless of the $\varepsilon_{A}$ values (red lines). On the other hand, the spin $\sigma=\,\dn$ case is presented in blue lines. In panels (a) and (b), the long-wire limit ($\varepsilon_{A}=0$) is addressed. In this limit, the current slope increases as the bias voltage increases, reaching the spin $\sigma=\,\up$ slope. Then, a half-integer differential conductance is obtained at zero bias. It is a direct consequence of the MZM leakage into the QD, as it was described for both the non-interacting \cite{liu2011,vernek2014} and the interacting case \cite{ruiz2015}. Thus, the leakage signature of the MZM is robustly observed even in the interplay with the Kondo resonance. On the other hand, within the short-wire limit, the current shows the slope breaking away from vanishing bias voltage [panel (c)], exhibiting a half-integer conductance at bias voltage $eV=2\varepsilon_{A}$, and the maximum resonance at zero bias voltage is restored [panel (d)]. In panel (e), we display the differential conductance at zero-bias as a function of $\ve_{A}$. From this panel, it is clear that the transition from half-integer to the approximate integer zero-bias conductance value is not abrupt, showing a gradual increment with $\ve_{A}$.

\begin{figure}[htbp]
\centering
\includegraphics[width=1.00\linewidth]{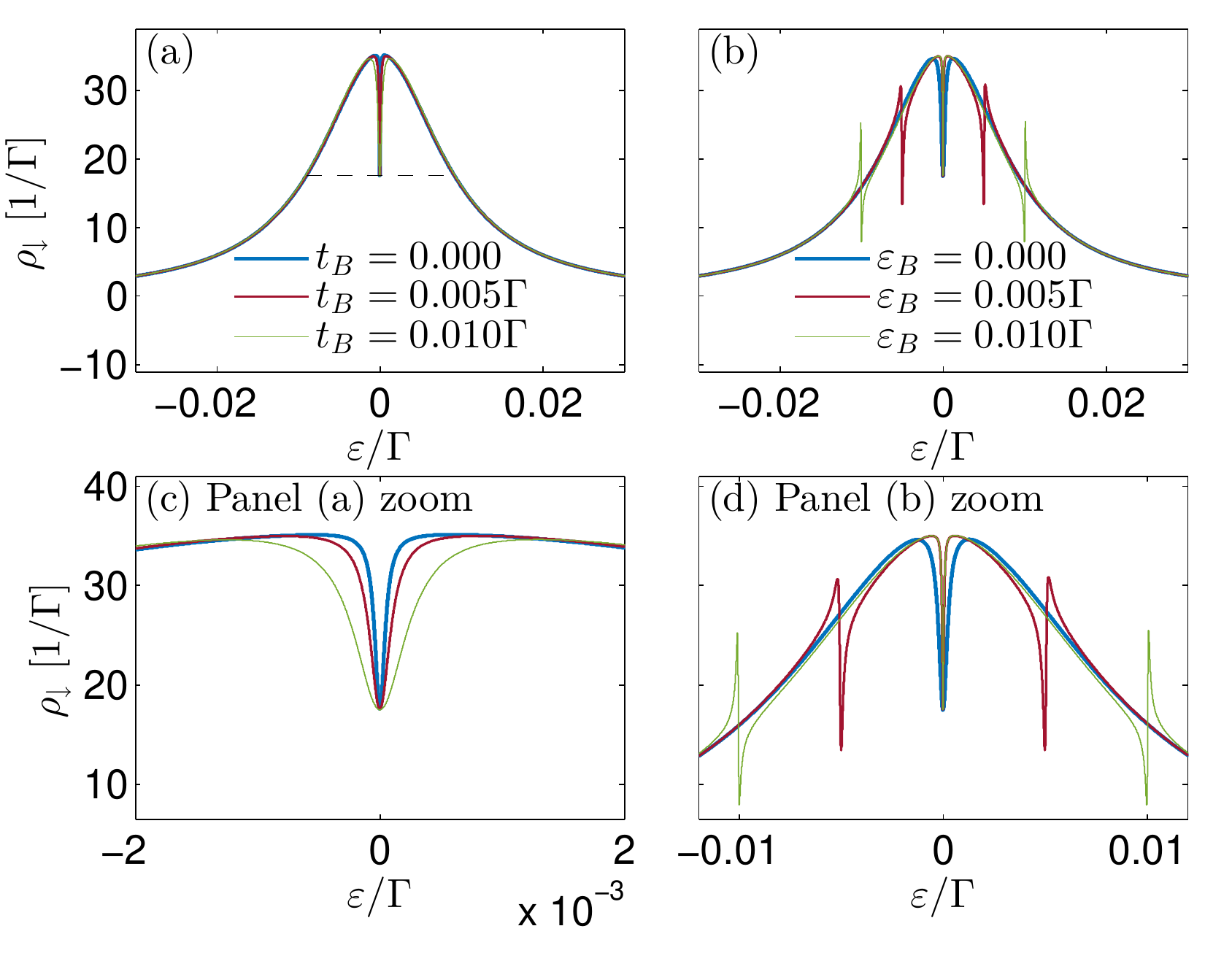}
\caption{QD's local density of states $\rho_{\dn}$ as function of the energy $\varepsilon$ for fixed $U=-10V_g$, $V_g=-10\Gamma$, $t_A=0.005\Gamma$ and $\varepsilon_A = \theta_A = \theta_B = 0$. Left panels [(a) \& (c)]: $\varepsilon_B=0$ and different $t_{B}$ values:  blue for $t_{B} = 0$, red for $t_{B} = 0.005\Gamma$ and green for $t_{B} = 0.010\Gamma$). The black dashed line represents the FWHM $0.564\Gamma$. Panel (c) is a zoom of panel (a). Right panels [(b) \& (d)]: $t_B=0.010\Gamma$ and different $\varepsilon_{B}$ values: blue for $\varepsilon_{B} = 0$, red for $\varepsilon_{B} = 0.005\Gamma$ and green for $\varepsilon_{B} = 0.010\Gamma$. Panel (d) is a zoom of panel (b).}
\label{rhod2}
\end{figure}

\begin{figure}[htbp]
\centering
\includegraphics[width=1.00\linewidth]{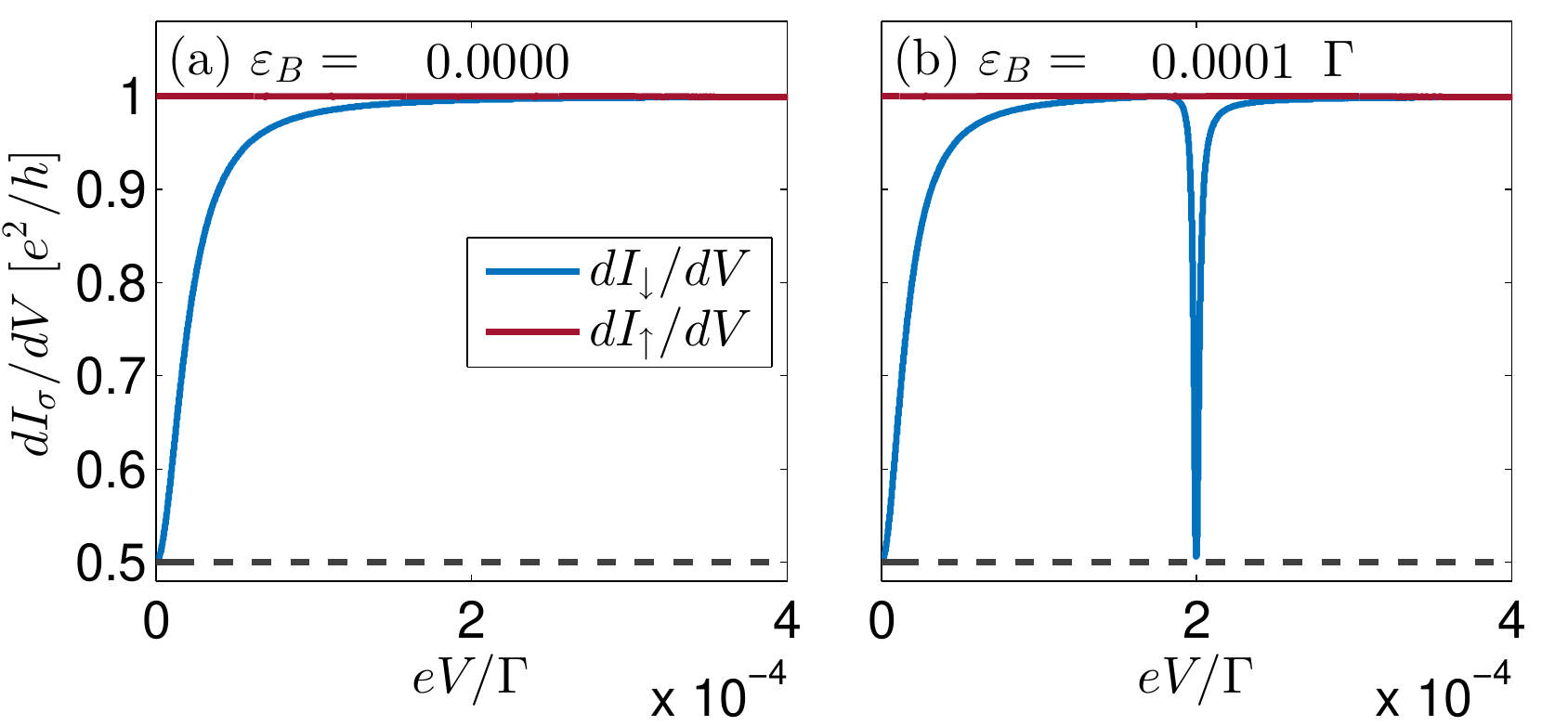}
\caption{Spin dependent differential conductance $dI_\sigma/dV$ as a function of the bias voltage $V$ for fixed $U=-10V_g$, $V_g=-10\Gamma$, $t_B=t_{A}/2=0.001\Gamma$, $\ve_{A}=0$ and $\theta_{A}=\theta_{B}=0$. Panel (a) is for $\varepsilon_B=0$; and panel (b) is for $\varepsilon_B=0.001\Gamma$. The black dashed lines represent the half-integer conductance.}

\label{IdGvsV_2}
\end{figure}

In what follows, we address the case with both TSCs connected. In Fig.\ \ref{rhod2} we present the QD's local density of states for the spin $\sigma=\,\dn$ component as function of the energy, considering both TSCs at the same phase $(\Delta\theta=0)$ and different $t_{B}$ values, while the $A$-TSC remains in the long wire limit
 ($\varepsilon_{A}=0$).
For both long wire limits [panels (a) and (c)], the resonance exhibits a dip, similar to the single wire case. Then, we interpret that both TSCs behave as an effective TSC, with a coupling strength $t_{\text{eff}}^{2}\propto t_{A}^{2}+t_{B}^{2}$. Whenever $\ve_{B}\neq 0$ [panels (b) and (d)], additional side interference profiles are observed at energies $\pm\ve_{B}$ (red and green lines), similar to the ones observed in Fig.\ \ref{rhod}, in addition to the one observed at $\ve=\ve_{A}=0$, which is due to the connection with the MZM from $A$-TSC.      

\begin{figure}[htbp]
\centering
\includegraphics[width=1.00\linewidth]{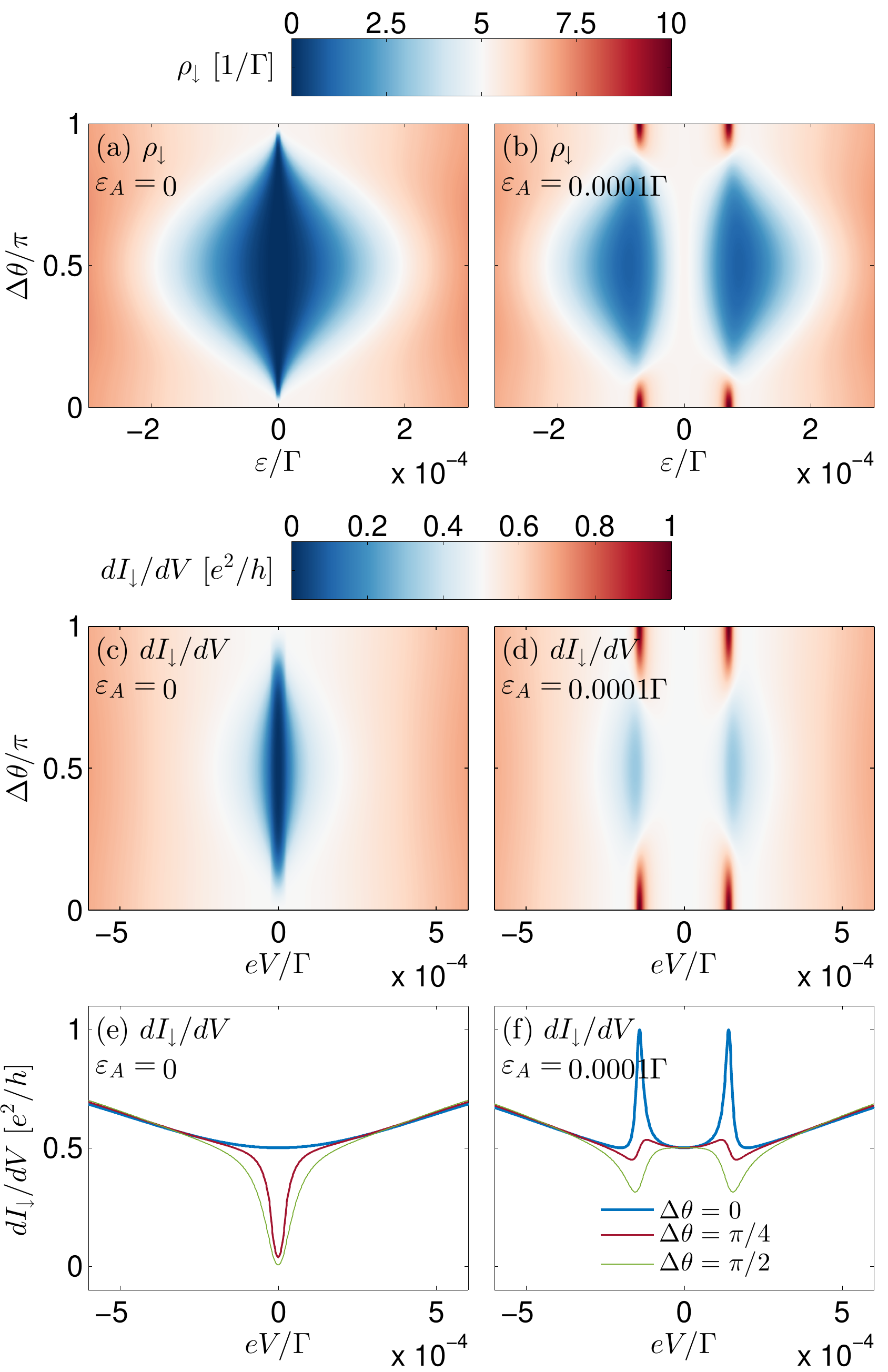}
\caption{Panels (a) and (b): QD's local density of states color maps for spin $\sigma=\,\dn$  as function of the energy and the phase difference. Panels (c) and (d): differential conductance $dI_\downarrow/dV$ color maps as function of the bias-voltage and the phase difference. For all panels we have $V_g=-10\Gamma$, $U=-10V_g$, $t_A=t_B=0.01\Gamma$ and $\varepsilon_B=0$. For the panels (a) and (c) we use $\varepsilon_A=0$ and for the panels (b) and (d) we use $\varepsilon_A=0.0001\Gamma$. In the panels (e) and (f) three particular cases of panels (c) and (d) are shown, for $\Delta\theta=0$, $\pi/4$ and $\pi/2$.}
\label{rhodGvsthA}
\end{figure}

In addition to the description above, the differential conductance is also affected, as we show in Fig.\ \ref{IdGvsV_2}, where the differential conductance is shown as a function of the bias-voltage, for a vanishing phase difference between both TSCs. In panel (a), since the MZMs belonging to the different wires do not interact destructively, the half-integer zero-bias conductance is still observed for the spin $\sigma=\,\dn$ component. On the other hand, in panel (b), for $\ve_{B}\neq 0$ an additional half-integer is obtained at $eV=2\ve_{B}$. Thus, the energy splitting of the MZMs belonging to the $B$-TSC (observed in the QD's local density of states) leads to understand that the half-integer behavior is not restricted to zero-bias only.    

We now address the role of the phase difference between both TSCs. In Fig.\ \ref{rhodGvsthA} we present QD's local density of states $\rho_{\dn}$ and differential conductance $dI_\downarrow/dV$ for fixed $\ve_{B}=0$. Panels (a) and (c) display $\rho_{\dn}$ and $dI_\downarrow/dV$ for $\ve_{A}=0$, respectively. From panel (a), an entire suppression of the resonance at $\ve=0$ whenever $\Delta\theta\neq n\pi$ (with $n=0,1,2,...$) is observed, and consequently panel (c) exhibits a vanishing differential conductance at zero-bias. This behavior can be explained as the destructive interference between MZMs placed in different wires, which interact through the QD and leads. It produces the weakening of the Kondo effect and the disappearance of MZMs signatures at zero energy point. 
 The Fano line shapes around zero bias can be characterized as:
 \begin{equation}
  \dfrac{dI_{\downarrow}}{dV} \approx \frac{e^2}{2h}\frac{\mid eV+q\mid^2}{(eV)^2+\gamma^2},  
 \end{equation}
where $q$ is the Fano parameter, $q=i q_0\sin{\Delta\theta}$, and $\gamma=\gamma_0\sin{\Delta\theta}$, the red and green lines in Fig. \ref{rhodGvsthA} panel (e) can be fitted with this expression, for the panel (f) a real part would have to be considered in $q$ and also change $eV\rightarrow eV\pm\varepsilon_A$. For the specific case of $\Delta\theta=\pi/2$, it is possible to interpret that the $\alpha$-type MZM from the wire $A$ became a $\beta$-type MZM from the $B$ wire (see Appendix\ \ref{apenA}). On the other hand, the case using $\ve_{A}\neq 0$ is presented in panels (b) and (d). At zero energy, the non vanishing dip in the QD's local density of states is still observed. Consequently, the zero-bias differential conductance exhibits a half-integer value, both attainable regardless of phase difference, since the MZMs from the $B$-TSC do not interfere with MZMs from $A$-TSC, due to their energy split. Nevertheless, the side resonances observed in QD's local density of states evolve to anti-resonances as the phase difference is tuned away from zero. The consequent evolution of the resonance regarding the phase difference can be interpreted and described as Fano-Majorana effect \cite{FanoMajorana2019RamosAndrade}. From the latter, it is remarkable how the interference phenomena behavior is robust in presence of electronic correlations.   

\section{Final remarks}\label{sec:conclu}

We studied a system formed by a single-level QD in the Kondo regime embedded between two TSCs, each hosting Majorana zero modes at their ends. We obtained the spin-resolved current and differential conductance across the QD provided by leads using the Green's function formalism to characterize the interplay between the Kondo effect and MZMs. The results obtained for spin $\sigma=\dn$ in the long wire limit show that the signature of the MZMs presence, i.e., half-integer differential conductance, prevails over signatures attributed to the Kondo effect at zero energy for one TSC connected and/or both TSCs connected with phase difference $\Delta\theta=n\pi$, with $n$ being an integer. For other phase values, destructive interference appears. Consequently, neither the Kondo effect nor signatures of MZMs are obtained since both the QD's local density of states and the differential conductance vanish at the exact zero energy point. Whenever one of the TSC connected is within the short wire limit, the zero bias half-integer conductance is obtained regardless of the phase difference. This feature is also present for other specific bias values related to the inter MZMs coupling energies. The coupling of MZMs does not influence the investigated quantities for spin $\sigma=\up$. Thus, our findings could be interpreted as a spin-resolved Kondo effect due to MZMs connection; whose obtained features can be accessed and measured in experiments.

\acknowledgments
J.P.R.-A is grateful for the funding of FONDECYT Postdoc. Grant No. 3190301 (2019). P.A.O. acknowledges support from FONDECYT Grant No. 1180914 and 1201876.

\onecolumngrid
\appendix

\section{Model for a topological superconductor coupled to a quantum dot} \label{apenA}

A quantum wire is modeled by a chain of $N$ sites, separated by a distance $a$, with a total length equal to $L$. The energy of an electron when it occupies one of these sites is $\varepsilon_w$. The coupling between neighboring sites, or hopping parameter, is $t_w$. Without any other interaction present, this model gives rise to an energy band having the dispersion relation $\varepsilon (k) = \varepsilon_w-2t_w \cos{(ka)}$.
We have considered that the quantum wire is exposed to a magnetic field and placed in proximity to a conventional BCS-type superconductor. Thus, it is possible to couple the spin-down band to the QD, but not possible for the spin-up band, which is located away from the QD's single-level due to the energy gap provided by the Zeeman effect. 
On the other hand, proximity to the conventional superconductor induces pairing of neighboring $p$-type electrons, producing a gap function $\Delta$, that has a smaller magnitude than in the BCS-type superconductor. 
It can be tuned such that it takes the same magnitude as the hopping parameter in the wire, $|\Delta| = t_w$. Additionally, setting the site energy of the electrons to zero, $\varepsilon_w = 0$, we have a half-full band of spin-down electrons; that is, we have the same number of electron-type and hole-type fermionic states. Thus, the Hamiltonian of this tuned system is expressed as
\begin{equation}
 H_{\text{TSC}} 
 =
-\sum_{\ell=1}^{N} |\Delta_{\ell} |
 \left\{ 
 \left( f_{\ell+1\, \downarrow}^{\dag} f_{\ell \downarrow} 
               +f_{\ell \downarrow}^{\dag} f_{\ell+1\, \downarrow} 
        \right)
+\left( e^{-2i\theta } f_{\ell\downarrow} f_{\ell+1\downarrow}
      +e^{2i\theta } f_{\ell+1\downarrow}^{\dag} 
f_{\ell\downarrow}^{\dag} 
  \right) 
\right\} \,,
\end{equation}
where $2\theta $ is the characteristic phase of the superconductor, and the $\ell$-index has the cyclic property $N + 1 = 1$.  
The gap function is $\Delta_{\ell} = |\Delta | e^{2i\theta }$ for $1 \leq \ell < N-1$, and for the hopping or pairing between the end sites of the chain is
$\Delta_{N} = |\Delta | e^{2i\theta } \, e^{-L/\xi} $, where $\xi$ is the coherence length. The Hamiltonian above can be diagonalized by transformations of the
Bogoliubov-Valatin type. To do this, let us first define Majorana fermions operators as follows:
\begin{equation}
 \alpha_{n} = \frac{1}{\sqrt{2}} 
 \left( e^{i\theta } {f}_{n\downarrow}^{\dag} 
       +e^{-i\theta } {f}_{n\downarrow}\right) 
 \,,\, \quad 
 \beta_{n}  = \frac{ -i }{\sqrt{2}} 
    \left( e^{i\theta } {f}_{n\downarrow}^{\dag} 
          -e^{-i\theta } {f}_{n\downarrow} \right) \,.
\end{equation}
Then, the Hamiltonian can be rewritten as
\begin{equation}
 H_{\text{TSC}} 
 =
\sum_{\ell=1}^{N} 2i |\Delta_{\ell}| \alpha_{\ell}\beta_{\ell+1}\,.
\end{equation}

In order to diagonalize, new fermionic operators are defined as

\begin{equation}
{c}_{\ell}^{\dag} 
 =
\frac{1}{\sqrt{2}} 
\left( {\alpha}_{\ell} -i{\beta}_{\ell+1} \right)
 \,,\quad \quad  
 {c}_{\ell}
 =
\frac{1}{\sqrt{2}}  
\left( {\alpha}_{\ell} +i{\beta}_{\ell+1} \right)
\, ; \quad 
\ell =1, \dots ,N.
\end{equation}
Considering these operators, the diagonalized Hamiltonian has the form
\begin{equation}
 H_{\text{TSC}} 
 =
 2|\Delta| \sum_{\ell=1}^{N-1} \left( c_{\ell}^{\dag}c_{\ell}-1 \right) 
 +
  2|\Delta|e^{-L/\xi} \left( c_{N}^{\dag}c_{N}-1 \right) \,.
\end{equation}

The weak coupling between the Majorana fermions at the ends of the TSC, $\alpha_{N}$ and $\beta_{1}$, is broken when the TSC interacts with the QD, and then we can write for the QD
\begin{equation}
 \alpha_{0} = \frac{1}{\sqrt{2}} 
 \left( e^{i\theta } {f}_{0\downarrow}^{\dag} 
       +e^{-i\theta } {f}_{0\downarrow}\right) 
 \,,\, \quad 
 \beta_{0}  = \frac{ -i }{\sqrt{2}} 
    \left( e^{i\theta } {f}_{0\downarrow}^{\dag} 
          -e^{-i\theta } {f}_{0\downarrow} \right)\text{,}
\end{equation}
allowing to write the TSC Hamiltonian and its coupling to the QD by means of the hopping $|t_{\nu}|$ as 
\begin{equation}
 {\cal H}_{\text{s-d}}  =
H_{\text{TSC}}
+ 2i |t_{\nu}| \alpha_{ 0} \beta_{ 1}
+ 2i |t_{\nu}| e^{-L/\xi}  \alpha_{N} \beta_{ 0}\text{.}
\end{equation} 
Rewriting the Hamiltonian, using usual fermionic operators, we obtain
\begin{equation}
\begin{aligned} 
 {\cal H}_{\text{s-d}}  & =
 2|\Delta|  \sum_{\ell = 1}^{N-1} 
 \left( c_{\ell}^{\dag} c_{\ell} -\tfrac{1}{2} \right)  
+2|\Delta|e^{-L/\xi} 
 \left( {c}_{N}^{\dag} {c}_{N} -\tfrac{1}{2} \right) 
\\ 
 & \; \; 
+|t_{\nu}| \left( 1+e^{-L/\xi} \right) 
 \left( e^{i\varphi_s} {f}_{0\downarrow}^{\dag} {c}_{N}^{\dag} 
       +e^{-i\varphi_s} {c}_{N} {f}_{0\downarrow} 
\right) 
\\ 
 & \; \; 
-|t_{\nu}|  \left( 1-e^{-L/\xi} \right) 
 \left( e^{i\varphi_s} {f}_{0\downarrow}^{\dag} {c}_{N}
       +e^{-i\varphi_s} {c}_{N}^{\dag} {f}_{0\downarrow} 
\right)\text{.}
\end{aligned}
\end{equation} 

In the equation above, the first term on the right hand side does not interact with the QD nor with the $N$-th fermion of the superconductor, thus this term can be ignored allowing us to obtain an effective Hamiltonian for the TSC and QD in the form
\begin{equation}
\begin{aligned} 
 {\cal H}_{\text{s-d,eff}}  & =
 2|\Delta| e^{-L/\xi}  {c}_{N}^{\dag} {c}_{N}
+|t_{\nu}|
 \left( e^{i\theta } {f}_{0\downarrow}^{\dag} {c}_{N}^{\dag} 
       +e^{-i\theta } {c}_{N} {f}_{0\downarrow} 
\right) 
-|t_{\nu}| 
 \left( e^{i\theta } {f}_{0\downarrow}^{\dag} {c}_{N}
       +e^{-i\theta } {c}_{N}^{\dag} {f}_{0\downarrow} 
\right)\text{.}
\end{aligned}
\end{equation} 

The latter description takes place for each TSC interacting with the QD. Each TSC will be identified by the index $\nu$ ($\nu = {A,B}$), thus we define
\begin{equation}
 \varepsilon_{\nu} =  2|\Delta_{\nu}| e^{-L_{\nu}/\xi_{\nu}}
 \,,\quad
 t_{\nu} = |t_{\nu}| e^{i\varphi_\nu}
 \,,\, \text{ and } \tilde{c}_{\nu} = c_N\text{.}
\end{equation}
Then, the effective Hamiltonian for one TSC denoted by index $\nu$ can be written out as
\begin{equation} \label{ham-TSC-QD}
 {\cal H}_{\nu d,\text{eff}}  =
 \varepsilon_{\nu} \tilde{c}_{\nu}^{\dag} \tilde{c}_{\nu}
+t_{\nu} {f}_{0\downarrow}^{\dag} \tilde{c}_{\nu}^{\dag} 
+t_{\nu}^*  \tilde{c}_{\nu} {f}_{0\downarrow} 
-t_{\nu}  {f}_{0\downarrow}^{\dag} \tilde{c}_{\nu}
-t_{\nu}^* \tilde{c}_{\nu}^{\dag} {f}_{0\downarrow}\text{.}
\end{equation}

\section{Treatment of the Coulomb interaction: Auxiliary bosons} \label{apenB}
In QDs the Coulomb interaction between electrons is appreciable and cannot be reduced to an effective field, so it is necessary to consider approximate methods. One of these uses a set of auxiliary bosons, as we will describe in what follows.

Let us consider an electronic system where there is a QD that has only one energy level $\varepsilon_d$, and where the Coulomb interaction, that arises in the double occupancy, is equal to $U$. Suppose that this QD has $n$ tunneling channels with coupling constants $t_{\ell}, \, \ell = 1, \dots , n$, to other parts of the system. Then, the Hamiltonian of such a system can be represented by
\begin{equation}
H = \sum_{\sigma=\{\uparrow,\downarrow\}} \varepsilon_d \hat{n}_{d,\sigma} 
 +U \hat{n}_{d,\uparrow} \hat{n}_{d,\downarrow}  
 +\sum_{\sigma=\{\uparrow,\downarrow\}} 
 \sum_{\ell =1}^{n} 
 \left( t_{\ell} c^{\dag}_{d \sigma} c_{\ell \sigma} +  t_{\ell}^*  c^{\dag}_{\ell \sigma} c_{d \sigma} \right)
 +H_{\text{others}}\,, 
\end{equation}
where $c_{d\sigma}^{\dag}$ ($c_{d\sigma}$) is the fermionic operator  to creation (annihilation) of an electron with spin $\sigma$ in the QD and $\hat{n}_{d\sigma} = c_{d\sigma}^{\dag} c_{d\sigma}$ is the number operator.

We considered the Coulomb interaction in the QD using auxiliary bosons, following the approach from Kotliar and Ruckenstein \cite{kotliar1986new}. The system is extended to a system that, in addition to the electronic system, includes four types of bosons associated with each of the four possible fermionic states of one QD. Then, $\hat{e}^{\dag}$($\hat{e}$) is the bosonic operator associated to creation(annihilation) of the empty state in the QD; $\hat{p}_{\sigma }^{\dag}$($\hat{p}_{\sigma}$) is the bosonic operator associated to creation(annihilation) of single state with spin $\sigma$ in the QD; and $\hat{d}^{\dag}$($\hat{d}$) is the bosonic operator associated to creation(annihilation) of doubly occupied state in the QD.

In the Hamiltonian of this extended system, the bosons are used to represent the Coulomb interaction, and the dynamic of the electrons is attached to the dynamic of the bosons. A restriction to fulfill on the  auxiliary  
bosons is that the number of each type of bosons must corresponds to the existence probability of the corresponding electronic state. Thus, the sum over the four bosonic numbers must be unity (completeness condition) 
\begin{equation}\label{restriction_completitude}
 |\hat{e}|^2 + |\hat{p}_{\uparrow }|^2  
 + |\hat{p}_{\downarrow}|^2 +|\hat{d}|^2  - 1 = 0\,.
\end{equation}
Besides, another restriction must be taken into account: count of electrons in electronic space should be equivalent to count in the auxiliary bosonic subspaces, then 
\begin{equation}  \label{restriction_ni}
 f_{0\, \sigma}^{\dag} f_{0\, \sigma} - |\hat{p}_{ \sigma }|^2 -|\hat{d}|^2
= 0\,.
\end{equation}

In the coupling terms, the fermionic annihilation operator of spin $\sigma$ for an electron in 
the QD, $f_{0\,\sigma}$, is attached to
equivalent operation over bosonic spaces, $\hat{Z}_{\sigma}$, as
\begin{equation}
 \hat{Z}_{\sigma} =  
 \hat{p}_{\bar{\sigma}}^{\dag} \hat{d} + 
        \hat{e}^{\dag} \hat{p}_{\sigma}\,.
\end{equation}
This operator on the boson spaces corresponds to two possible transitions associated with the destruction of the electron. Then, using the restrictions given by Eqs.\ (\ref{restriction_completitude}) and (\ref{restriction_ni}) with the Lagrange's multiplier $\lambda_{0}^{(1)}$ and $\lambda_{0\, 
\sigma}^{(2)}$, respectively, we obtain the following effective Hamiltonian

\begin{equation}
 \begin{aligned} 
  H_{\text{eff}} = &
 = \sum_{\sigma=\{\uparrow,\downarrow\}} \varepsilon_d \hat{n}_{d,\sigma} 
 +\sum_{\sigma=\{\uparrow,\downarrow\}} 
 \sum_{\ell =1}^{n} 
 \left( t_{\ell} \hat{Z}_{\sigma}^{\dag} c^{\dag}_{d \sigma} c_{\ell \sigma} +  t_{\ell}^* \hat{Z}_{\sigma} c^{\dag}_{\ell \sigma} c_{d \sigma} \right)
 +H_{\text{others}} 
 \\ & 
+\lambda_{0}^{(1)}
 \left(  |\hat{e}|^2 + |\hat{p}_{\uparrow }|^2  
 + |\hat{p}_{\downarrow}|^2 +|\hat{d}|^2  - 1 \right)  
 +\sum_{\sigma } \lambda_{0\, \sigma}^{(2)}
 \left( \hat{n}_{0\,\sigma} - |\hat{p}_{ \sigma }|^2 
-|\hat{d}|^2 \right)
 +U |\hat{d}|^2  \,,
 \end{aligned}
\end{equation}
where the Coulomb interaction between electrons that has been replaced by its equivalent in boson space
 associated with double occupancy.
Now, in a mean-field approximation, the dynamics of electrons are considered to react only to the mean values of the auxiliary bosons, and real numbers can replace the mean values of the boson number operators.
Hereafter, the notation $b$ means the average value of the bosonic operator
$\hat{b}$, i.e. 
\[
 b = \left\langle \hat{b}^{\dag} \right\rangle 
        = \left\langle \hat{b} \right\rangle 
        \,, \; b=\left\{ e,\, p_{\uparrow},\, p_{\downarrow}, \,d \right\}\,.
\]

 Kotliar and Ruckenstein \cite{kotliar1986new} care about to get the correct $U=0$ limit in the saddle-point approximation, thus the operator $\hat{Z}_{\sigma}$ is replaced by the mean value 
\[
 \hat{Z}_{\sigma} \to Z_{\sigma} =  
 \frac{ p_{\sigma} e +d p_{\bar{\sigma}}  } 
     { \sqrt{1-p_{\sigma}^2 - d^2 } 
       \sqrt{1-e^2-p_{\bar{\sigma}}^2  }  }\,.
\]
Unlike them, we are interested in obtaining the correct behavior for finite $U$ of the non-degenerate Anderson model, in which the Kondo effect appears and with a Kondo temperature that follows a well-established relationship. For this we replace the operators $\hat{Z}_{\sigma}$ by the mean values

\begin{equation}
  \hat{Z}_{\sigma} \to Z_{\sigma} =  
 \left( p_{\sigma} e +d p_{\bar{\sigma}} \right) 
 \, F_Z(U,\varepsilon_d)\,,    
\end{equation}
where the function $F_Z$, which depends on the Coulomb interaction $U$ and on the energy level $\varepsilon_d$, has a value defined such that in the limit of Anderson non-degenerate, the half-width of the resonance peak at zero temperature is equal to the established Kondo temperature \cite{KondoTemperature,pruschke1989anderson,tsvelick1983exact}, whose expression is given by

\begin{equation}
 k_B T_K = \frac{\text{min}(U,W)}{2\pi} \sqrt{I} e^{-\pi/I}
 \,, \quad 
  I = 2\Gamma_{\text{eff}} 
 \left[ \frac{1}{|\varepsilon_d|} +\frac{1}{U+\varepsilon_d} \right] \,,
\end{equation}
where $\Gamma_{\text{eff}} = 2 \Gamma$ and $W$ is the half width of the conducting bands coupled to the QD. Then, within the mean field approximation (mfa), the Hamiltonian is given by
\begin{equation}
 \begin{aligned} 
  H_{\text{mfa}} = &
 \sum_{\sigma=\{\uparrow,\downarrow\}} \tilde{\varepsilon}_{\sigma} \hat{n}_{d,\sigma} 
 +\sum_{\sigma=\{\uparrow,\downarrow\}} 
 \sum_{\ell =1}^{n} 
 \left( \tilde{t}_{\ell \sigma} c^{\dag}_{d \sigma} c_{\ell \sigma} +     \tilde{t}_{\ell \sigma}^* c^{\dag}_{\ell \sigma} c_{d \sigma} \right)
 +H_{\text{others}} 
 \\ & 
+\lambda_{0}^{(1)}
 \left( e^2 + p_{\uparrow }^2  
 + p_{\downarrow}^2 +d^2  - 1 \right)  
 -\sum_{\sigma } \lambda_{0\, \sigma}^{(2)}
 \left( p_{ \sigma }^2 +d^2 \right)  +U |\hat{d}|^2  \,,
 \end{aligned}
\end{equation}
where $\tilde{t}_{\ell \sigma} = t_{\ell} Z_{\sigma}$,  and 
$\tilde{\varepsilon}_{\sigma} = 
 \varepsilon_{d} + \lambda_{0\, \sigma}^{(2)}$.

The parameters $ e$, $ d$, $ p_{\sigma}$, 
$ \lambda^{(1)}_{0}$, and $\lambda^{(2)}_{0\, \sigma}$ can be determined by minimizing the ground state energy of the Hamiltonian 
$H_{\text{mfa}}$. 
Conditions for minimal energy 
together with the application of the Hellmann-Feynman theorem, 
$\partial\langle H\rangle/\partial x  = 
\langle\partial H/\partial x\rangle$, gives a set of equations that can be solved self-consistently.

\twocolumngrid

\bibliographystyle{apsrev4-1}
\bibliography{biblio3}

\end{document}